\documentclass[twocolumn,showpacs,amssymb,amsmath,aps,pra]{revtex4}
\newcommand{\half}{\mbox{$\frac{1}{2}$}}
\usepackage{graphics}
\usepackage{bm}
\begin{document}
\title{Global entanglement in $XXZ$ chains}
\author{N. Canosa, R. Rossignoli}
\affiliation{Departamento de F\'{\i}sica-IFLP, Facultad de Ciencias Exactas, 	
Universidad Nacional de La Plata, C.C.67, La Plata (1900), Argentina}
\begin{abstract}
We examine the thermal entanglement of $XXZ$-type Heisenberg chains in the
presence of a uniform magnetic field along the $z$ axes, through the evaluation 
of the negativity associated with bipartitions of the whole system and
subsystems. Limit temperatures for non-zero global negativities are shown to
depend on the asymmetry $\Delta$ but not on the uniform field, and can be much
higher than those limiting pairwise entanglement. It is also shown that global
bipartite entanglement may exist for $T>0$ even for $\Delta\geq 1$, i.e., when
the system is fully aligned (and hence separable) at $T=0$, and that the
bipartition leading to the highest limit temperature depends on $\Delta$.
\end{abstract}
\maketitle
\section{Introduction}
Quantum entanglement is a fundamental trait of quantum mechanics, representing
the non-local correlations with no classical analogue that can be exhibited by
composite quantum systems. Interest on entanglement has been triggered in the
last few years by the discovery of its potential for developing new forms of
information transmission and processing \cite{Be.93,Ek.91,Di.95,Be.00}, being a
resource in Quantum Information science \cite{NC.00}. Entanglement has also
provided a new perspective for the analysis of correlations and transitions in
many-body quantum systems \cite{ON.02,OS.02,V.03,VW.04,T.04}. In systems at
finite temperature $T$, two fundamental questions which immediately arise
\cite{N.98,Ar.01} are: a) the determination of the limit temperatures for the
existence of different kinds of entanglement and b) the possible emergence of
entanglement for $T>0$ when the ground state is separable due to entangled
excited states.

Let us first recall that a mixed state $\rho$ of a bipartite quantum system
$A+B$ is said to be {\it separable} or {\it classically correlated} if it can
be expressed as a convex combination of product densities \cite{W.89}, i.e.,
$\rho=\sum_{\alpha}q_\alpha\rho_A^\alpha\otimes \rho_B^\alpha$, with
$q_\alpha>0$, $\sum_\alpha q_\alpha=1$ and $\rho_{A}^\alpha$, $\rho_B^\alpha$
density matrices for each component. Otherwise, $\rho$ is {\it entangled} or
{\it inseparable}. Separable states can be created by local operations and
classical communication (LOCC) and satisfy Bell inequalities as well as other
classical properties such as being more disordered globally than locally
\cite{NK.01,RC.02}. Pure states ($\rho^2=\rho$) are separable if and only if
$\rho=\rho_A\otimes\rho_B$, but this is not necessarily the case for mixed
states, where the determination of separability is in general an NP hard
problem \cite{Gu.03,DPS.04}. Accordingly, while for pure states bipartite
entanglement can be measured in terms of the entropy of the reduced density of
a subsystem \cite{Be.96}, a rigorous computable measure of entanglement for
mixed states has so far been obtained just for the case of two qubits, through
the concurrence \cite{W.98}.

Nonetheless, bounds for entanglement in general mixed states can be obtained by
means of the {\it negativity} \cite{VW.02,ZHSL.98,HHH.00}, which is a measure
of the degree of violation of the criterion of positive partial transpose (PPT)
\cite{P.96,HHH.96} in entangled states and is easily computable. Although the
PPT criterion is a necessary separability condition, sufficient just for
two-qubit or qubit+qutrit systems, the negativity fulfills some fundamental
properties of an entanglement measure \cite{VW.02}, being an entanglement
monotone and providing bounds to the teleportation capacity and distillation
rate. In $n$-qubit systems at finite temperature, it can then be employed
\cite{RC.05} to detect the entanglement between the components of any
bipartition $\{m\}-\{n-m\}$ of the {\it whole} system, as well as of {\it any}
subsystem, beyond the level of pairwise two-qubit entanglement measured by the
concurrence. Another fundamental result is that any state $\rho$ is completely
separable (convex combination of $n$-product densities) if it is sufficiently
close to the fully mixed state $I/d$ \cite{ZHSL.98,B.99,GB.02} ($d$ is the
system dimension). This ensures that any canonical thermal state
$\rho(T)\propto \exp[-H/T]$ of a finite system described by a Hamiltonian $H$
becomes {\it separable} above a  {\it finite} limit temperature, since it will
be as closed as desired to $I/d$ for sufficiently high $T$. In particular, no
bipartite entanglement can arise if ${\rm Tr}(\rho-I/d)^2\leq [d(d-1)]^{-1}$
\cite{GB.02}.

The aim of this work is then to examine, by means of the negativity, the {\it 
global} bipartite entanglement of thermal states $\rho(T)$ of $n$-spin chains
interacting through an $XXZ$-type coupling placed in a transverse magnetic
field. Heisenberg chains can be employed for solid state quantum computers
\cite{LDV.98} and $XXZ$ chains have been used to describe quantum computers
based on NMR \cite{GBF.03} and on electrons on Helium \cite{PD.99,S.03}.
Relevant studies of the {\it pairwise} thermal entanglement between two qubits
in Heisenberg chains have been made
\cite{ON.02,Ar.01,GBF.03,G.01,W.02,WFS.01,KS.02,CR.04}, revealing rich
phenomena. In small $XXZ$ chains \cite{WFS.01,GBF.03} it has been shown that
there is no pairwise entanglement for $T\geq 0$ for anisotropies $\Delta>1$,
i.e., when the ground state is fully aligned for any field, and that the
corresponding limit temperature remains bounded for large negative $\Delta$ if
$n$ is odd (vanishing $\forall$ $\Delta$ in the $n=3$ antiferromagnetic case at 
zero field \cite{WFS.01}). In contrast, we will show here that global 
entanglement between $m$ and $n-m$ qubits {\it may also exist for $\Delta>1$ if 
$n\geq 4$ and $T>0$}, implying a {\it thermal reentry}. Moreover, limit 
temperatures for non-zero global negativities {\it do not saturate} for 
$\Delta\rightarrow-\infty$ in odd systems, but diverge in {\it both} the ferro-
and antiferromagnetic cases, even for $n=3$. These temperatures are {\it
independent} of the magnetic field but dependent on $\Delta$ and the
bipartition, displaying crossings as $\Delta$ is varied. Numerical results up
to $n=10$ qubits will be provided, together with full analytical results for
$n=3$.

\section{Formalism}
We will consider an $XXZ$ Hamiltonian for $n$ qubits or spins of the form
\begin{eqnarray}
H&=&bS_z-\sum_{i<j}[v_x^{ij}(s_x^is_x^{j}+s_y^{i}s_y^{j})+v_z^{ij}s_z^is_z^{j}]
\label{H1}
\end{eqnarray}
where $\mathbf{s}_i$ denotes the spin operator at site $i$, $S_z=\sum_i s_z^i$
is the total spin $z$-component, $b$ accounts for the Zeeman coupling to a
uniform magnetic field and
$s_x^is_x^{j}+s_y^{i}s_y^{j}=(s_+^is_-^j+s_-^is_+^j)/2$ is the hopping or
entangling term. Our attention will be centered on a cyclic chain with nearest
neighbor coupling ($v_\alpha^{ij}=v_\alpha\delta_{j,i+1}$ for $\alpha=x,z$,
with $n+1\equiv 1$), although some results for the fully connected case
($v_\alpha^{ij}=v_\alpha$ $\forall$ $i<j$) will also be commented for
comparison. In the first case the spectrum of $H$ and the entanglement of its
eigenstates are independent of the sign of $v_x$ for even $n$, as it can be
changed by a rotation around the $z$ axes at odd sites ($s_{x,y}^i\rightarrow
(-1)^is_{x,y}^i$). In any case, $[H,S_z]=0$, so that the eigenstates of $H$
have good quantum number $M$ (eigenvalues of $S_z$).

We will consider the $n$-spin thermal state
\begin{equation}
\rho(T)=Z^{-1}\exp[-H/T]\,,\label{rt}
\end{equation}
where $Z={\rm Tr}\exp[-H/T]$ is the partition function and $T>0$ the
temperature (we set Boltzmann constant $k=1$). Global entanglement between $m$
and $n-m$ selected qubits will be analyzed by means of the negativity
\cite{VW.02}
\begin{equation}
N_p[\rho]=\half{\rm Tr}(|\rho^{t_p}|-1)\,,\label{N1}
\end{equation}
where $p\equiv\{m\}-\{n-m\}$ denotes the bipartition and $\rho^{t_p}$ the
ensuing partial transpose of $\rho$ \cite{P.96}. Eq.\ (\ref{N1}) is just the
absolute value of the sum of the negative eigenvalues of $\rho^{t_p}$ (as ${\rm
Tr}\rho^{t_p}=1$), so that according to the PPT criterion \cite{P.96,HHH.96},
$N_p[\rho]>0$ indicates entanglement between the $m$ selected qubits and the
rest. Eq.\ (\ref{N1}) satisfies as well properties of an entanglement measure
\cite{VW.02}, being a {\it convex} function of $\rho$ and an entanglement
monotone (it does not increase under LOCC). Moreover, distillable entangled
states satisfy $N_p[\rho]>0$ \cite{HHH.00}. Although the present analysis
leaves away bound PPT entangled states and does not capture all aspects of
$n$-partite entanglement (separability of all bipartitions does not necessarily
imply full separability), it goes beyond the standard analysis based on
pairwise two-qubit entanglement (the latter may vanish even if $N_p[\rho]>0$
for all global bipartitions, as occurs in GHZ states \cite{RC.05}). For {\it
pure} states $\rho=|\Psi\rangle\langle\Psi|$, it can be shown, by means of the
Schmidt decomposition \cite{NC.00} $|\Psi\rangle=\sum_\alpha
\sqrt{\lambda^p_\alpha}|\alpha_{\{m\}}\rangle|\alpha_{\{n-m\}}\rangle$, that
\begin{equation}
N_p[\rho]=\sum_{\alpha<\alpha'}\sqrt{\lambda^p_\alpha\lambda^p_{\alpha'}}
=\half[(\sum_\alpha\sqrt{\lambda^p_\alpha})^2-1]\;\;(\rho^2=\rho)\,,\label{N0}
\end{equation}
where  $\lambda^p_\alpha$ represent the eigenvalues of the reduced density
$\rho_{\{m\}}\equiv{\rm Tr}_{\{n-m\}}\rho$ of the $m$ selected qubits (the same
as those of $\rho_{\{n-m\}}$ when $\rho^2=\rho$). Eq.\ (\ref{N0}) differs from
the entanglement of formation \cite{Be.96}
$E[\rho]\propto-\sum_{\alpha}\lambda^p_\alpha\log_2\lambda^p_\alpha$, but is as
well a measure of the disorder of the reduced system (see Appendix), satisfying
$N_p[\rho]\geq N_p[\rho']$ if $\lambda^p\prec {\lambda'}^p$, where $\prec$
denotes ``majorized by'' (or ``more mixed than'') \cite{NC.00}:
$\sum_{\alpha=1}^k \lambda^p_{\alpha}\leq
\sum_{\alpha=1}^k{\lambda'}^p_{\alpha}$ for $k=1,\ldots,d_m$, where
$\lambda_\alpha^p$, ${\lambda'}_\alpha^p$ are sorted in decreasing order and
$d_m=2^m$ is the subsystem dimension ($m\leq n/2$). The formal maximum is then
obtained for a uniform distribution $\lambda^p_\alpha=1/d_m$, in which case
$N_p[\rho]=(d_m-1)/2$. Due to convexity, this is also the maximum for non-pure
states.

In the nearest neighbor cyclic chain, global negativities $N_p[\rho]$ depend on
the number $m$ of selected qubits and on their spacings. For instance, in a
3-qubit chain $abc$ there is single distinct global negativity $N_{a-bc}$
($=N_{b-ac}=N_{c-ba}$),  whereas for $n=4$ qubits $abcd$, there are three
distinct negativities $N_{a-bcd}$, $N_{ab-cd}$ and $N_{ac-bd}$, which measure
respectively the entanglement between:  one qubit and the rest, adjacent pairs,
non-adjacent pairs. Each negativity $N_p[\rho(T)]$ will have its {\it own
limit temperature} $T_p$ above which it vanishes, although the behavior for
$T<T_p$ may not be monotonous (even entanglement vanishing plus reentry may
occur for $T<T_p$ \cite{KS.02,CR.04}). A remarkable feature of these global
limit temperatures is that for Hamiltonians of the form (\ref{H1}), they are
strictly {\it independent} of the uniform field $b$ \cite{RC.05} (see
Appendix), even though $N_p[\rho(T)]$, as well as the entanglement of the
ground state, depend of course on $b$.

The bipartite entanglement of subsystems of $m<n$ qubits can be analyzed in a
similar way by evaluation of the reduced negativities $N_{p'}[\rho_{\{m\}}]$
determined by the reduced density $\rho_{\{m\}}$, with $p'=\{k\}-\{m-k\}$ a
bipartition of the subsystem. These negativities will in general also depend on
the relative location of the $k$ qubits, and satisfy inequalities of the form
\cite{VW.02} $N_{\{k\}-\{m-k\}}[\rho_{\{m\}}]\leq
N_{\{k\}-\{m+1-k\}}[\rho_{\{m+1\}}]$, as tracing out one qubit is a LOCC
operation. The associated limit temperatures will then satisfy
$T_{\{k\}-\{m-k\}}\leq T_{\{k\}-\{m+1-k\}}$, being thus higher for global
bipartitions. In particular, the negativity of one qubit with the rest
($N_{a-*}$) is an upper bound to all pairwise negativities $N_{a-b}$,
$N_{a-c}$, etc., so that $T_{a-*}$ is an upper bound to all pairwise limit
temperatures. Limit temperatures $T_{p'}$ of reduced negativities
$N_{p'}[\rho_{\{m\}}(T)]$ {\it depend} in general on the field $b$.

For instance, for a three-qubit cyclic chain there is a single reduced
two-qubit density $\rho_{ab}={\rm Tr}_c\rho$ (any  other choice of pair is
equivalent), with negativity $N_{a-b}[\rho_{ab}]\leq N_{a-bc}[\rho]$, while for
a four qubit chain  there is one distinct three-qubit density $\rho_{abc}={\rm
Tr}_{d}\,\rho$, with {\it two} different negativities $N_{a-bc}[\rho_{abc}]$,
$N_{b-ac}[\rho_{abc}]$, satisfying $N_{a-bc}\leq (N_{a-bcd},N_{ab-cd})$,
$N_{b-ac}\leq (N_{a-bcd},N_{ac-bd})$. There are also two different pair
densities $\rho_{ab}={\rm Tr}_{cd}\rho$, $\rho_{ac}={\rm Tr}_{bd}\rho$, whose
negativities $N_{a-b}$, $N_{a-c}$, measure the entanglement of contiguous and
non-contiguous qubits and satisfy $N_{a-b}\leq (N_{a-bc},N_{b-ac})$,
$N_{a-c}\leq N_{a-bc}$.

\section{Results}
We consider in what follows a cyclic chain with nearest neighbor coupling, and
define the anisotropy as $\Delta\equiv v_z/|v|$, with $v\equiv v_x$. At $T=0$
and fixed $b\neq 0$, the ground state of the Hamiltonian will experience a
series of $[n/2]$ transitions $|M|\rightarrow |M|+1$ as $\Delta$ increases,
starting from $|M|=0$ ($1/2$) for $n$ even (odd) and large negative $\Delta$,
and ending in an {\it aligned state} with maximum spin $|M|=n/2$ for
$\Delta>\Delta_c(b)$. This state is completely separable while all ground
states with $|M|<n/2$ are entangled,  so that $\Delta_c(b)$ indicates the
entangled-separable border at $T=0$. The energy of the aligned state
$|\Psi_0\rangle$ ($=|\!\!\downarrow\ldots \downarrow\rangle$ for $b>0$) is
$E_{0}=-n(2|b|+v_z)/4$, while for $v>0$, that of the lowest $|M|=1$ state,
which is a $W$-type state
($|\Psi_1\rangle\propto\sum_{i=1}^ns_+^i|\Psi_0\rangle$ for $b>0$), is
$E_{1}=E_{0}+|b|+v_z-v$, so that the transition occurs at
\begin{equation}
\Delta_c(b)=1-|b/v|\;\;\;\;\;(v>0\;{\rm if}\;n\;{\rm odd}) \,.\label{dc}
\end{equation}
Eq.\ (\ref{dc}) is of course also valid for $v<0$ if $n$ is even.

The entanglement borders for $T>0$ are, however, quite different. As limit
temperatures for non-zero global negativities are independent of the field $b$,
so will be the concomitant borders $\Delta^p_c(T)$. Global bipartite
entanglement for $T>0$ and $b\neq 0$ will then also arise for
$\Delta_c(b)<\Delta<1$. Moreover, we will see that it may also arise for
$\Delta\geq 1$ if $n\geq 4$.

{\it Two and three qubit case}. Let us first discuss the behavior for two and
three qubits, where analytical expressions for $T>0$ can be found (for $n=2$ we
consider just a single term $v_\alpha^{12}$ in (\ref{H1})). $H$ can be
rewritten in these cases in terms of the total spin components
$S_\alpha=\sum_{i=1}^n s^i_\alpha$ as
\begin{equation}
H=bS_z-\half [v(S_x^2+S_y^2)+v_zS_z^2]+E_0\,,\label{HS}
\end{equation}
where $E_0=n(2v+v_z)/8$, so that its eigenstates have good total spin $S$ with
energies
\begin{equation}
E^S_M=bM-\half[vS(S+1)+M^2(v_z-v)]+E_0 \,,\label{ESM}
\end{equation}
where $S=0,1$, for $n=2$, and $S=1/2$ (two-fold degenerate), $3/2$, for $n=3$,
with $|M|\leq S$. In the fully connected case Eqs.\ (\ref{HS})-(\ref{ESM}) are
of course valid for all $n$, with $S=0\,(1/2),\ldots,n/2$ for $n$ even (odd).

\begin{figure}[t]
\vspace*{0.cm}

 \centerline{\scalebox{0.5}{\includegraphics{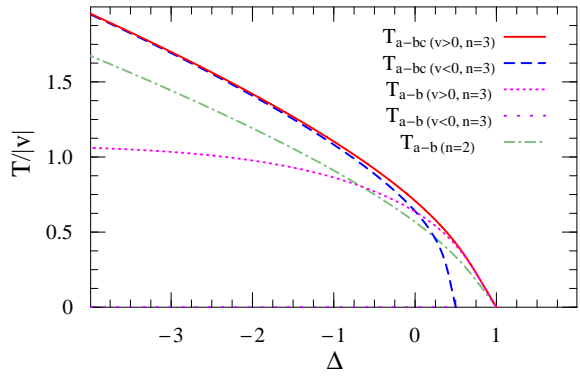}}}
 \vspace*{-.25cm}

\caption{(Color online). Limit temperatures $T_{a-bc}$ for global bipartite
entanglement for $n=3$ qubits as a function of the asymmetry $\Delta=v_z/|v|$,
for both signs of $v$ (they are independent of the uniform field $b$). Also
depicted are the corresponding limit temperatures for pairwise entanglement
$T_{a-b}$ for $b=0$ (that for  $v<0$ vanishes), and the concomitant result for
$n=2$ qubits (the same for both signs of $v$ and any $b$).}
 \label{f1}\vspace*{-.25cm}
\end{figure}

For $n=2$, the eigenstates of $H$ are then either separable (the aligned states
($|SM\rangle=|1,\pm 1\rangle$) or maximally entangled (the Bell states
$|SM\rangle=|10\rangle,|00\rangle\propto
|\!\!\uparrow\downarrow\rangle\pm|\!\!\downarrow\uparrow\rangle$), so that the
ground state is maximally entangled for $\Delta<\Delta_c(b)$ (with $S=1$ if
$v>0$ and $S=0$ if $v<0$). On the other hand, any two-qubit mixed state of the
form $\rho=\sum_{S,M}p^S_M|SM\rangle\langle SM|$ is entangled whenever
\begin{equation}
 |p^1_0-p^0_0|>2\sqrt{p^1_1 p^1_{-1}}\,, \label{EP}
\end{equation}
(see Appendix), which in the thermal case (\ref{rt}) ($p^S_M=e^{-E^S_{M}/T}/Z$)
leads to the $b$-independent border
\begin{equation}
\Delta\leq 1-2t\ln[2/(1-e^{-1/t})]\,,\;\;\;t\equiv T/|v|\,.\label{EV}
\end{equation}
Thermal entanglement arises then for $\Delta<1$ $\forall$ $b$, implying reentry
for $T>0$ if $\Delta_c(b)<\Delta<1$. The limit temperature $T_{a-b}(\Delta)$
determined by (\ref{EV}) is a {\it decreasing} function of $\Delta$ (see Fig.\
\ref{f1}), that vanishes for $\Delta\rightarrow 1$ and diverges for
$\Delta\rightarrow-\infty$ ($\Delta\leq 1-2t\ln(2t)$ for $t\gg 1$) .

For $n=3$ qubits, the behavior of {\it global} bipartite entanglement is
qualitatively similar to that for $n=2$ (Fig.\ 1). For $v>0$, the ground state
is a $W$-type entangled state if $\Delta<\Delta_c(b)$ \cite{B.05}
($|SM\rangle=|\frac{3}{2},-\frac{1}{2}\rangle\propto
|\!\!\uparrow\downarrow\downarrow\rangle+
|\!\!\downarrow\uparrow\downarrow\rangle+
|\!\!\downarrow\downarrow\uparrow\rangle$ for $b>0$), with global negativity
$N_{a-bc}=\sqrt{2}/3 \approx 0.47$ and pair negativity
$N_{a-b}=(1-\sqrt{5})/6\approx 0.21$, becoming the aligned state
$|\frac{3}{2},\pm\frac{3}{2}\rangle$ for $\Delta>\Delta_c(b)$. Let us add that
if $b=0$, the states $|\frac{3}{2},\pm \frac{1}{2}\rangle$ become degenerate
but their mixture $\half\sum_{M=\pm 1/2}|\frac{3}{2}M\rangle\langle
\frac{3}{2}M|$ (the $T\rightarrow 0$ limit of $\rho(T)$ for
$\Delta<\Delta_c(b)$) remains entangled,  with lower values
$N_{a-bc}=(\sqrt{3}-1)/3\approx 0.24$, $N_{a-b}=1/6$.

On the other hand, for any 3-qubit mixed state $\rho=\sum_{S,M}p^S_M P^S_M$,
where $P^S_M$ denotes the projector onto the subspace with total spin $S$ and
$z$-component $M$, the global negativity $N_{a-bc}$ will be non-zero if and
only if (see Appendix)
\begin{equation}
|p^{3/2}_{\nu/2}-p^{1/2}_{\nu/2}|>\sqrt{3p^{3/2}_{3\nu/2}
(p^{3/2}_{-\nu/2}+p^{1/2}_{-\nu/2}/2)}\,,\label{e3r}
\end{equation}
for $\nu=1$ {\it or} $\nu=-1$, which in the thermal case implies
\begin{equation}
\Delta<1-t\ln[\frac{3(2+e^{-3/2t})}{2(1-e^{-3/2t})^2}]\;\;\;(v>0)\,.
\label{e4r}
\end{equation}
Global entanglement is then again feasible for $\Delta<1$ $\forall$ $b$
($\Delta<1-t\ln 3$ for $t\ll 1$) and the ensuing limit temperature $T_{a-bc}$
is a decreasing function of $\Delta$, that vanishes for $\Delta\rightarrow 1$
and {\it diverges} for $\Delta\rightarrow-\infty$ ($\Delta<1-2t\ln(\sqrt{2}t)$
if $t\gg 1$) (see Fig.\ \ref{f1}).

The behavior of the {\it pairwise} limit temperature $T_{a-b}$ for $n=3$ is
however quite different \cite{WFS.01,GBF.03}. The reduced two-qubit density
$\rho_{a-b}$ is entangled in a smaller region which {\it depends} on the field
$b$ and is determined by the equation (see Appendix)
\begin{equation}
\Delta\leq
1-t\ln\frac{3}{\sqrt{\gamma^2(\eta^2-1)+2(1+\eta)(1-\alpha)^2}-\gamma\eta}\,,
 \label{e6r}\end{equation}
where $\gamma=1+2\alpha$, $\alpha=e^{-3/2t}$ and $\eta=\cosh(b/T)$. For $t\ll
1$, $\Delta\alt 1-t\ln 3$ as before, so that for $T>0$, $\rho_{a-b}$ is also
entangled for $\Delta<1$ $\forall$ $b$. However, the denominator in (\ref{e6r})
{\it vanishes} at a {\it finite} temperature $t_c(b)$, implying that
$T_{a-b}/v$ approaches a {\it finite} limit $t_c(b)$ for
$\Delta\rightarrow-\infty$, in contrast with $T_{a-bc}$. Hence, for large
negative $\Delta$ there is a large range of temperatures where only {\it
global} bipartite entanglement persists, with
$T_{a-bc}/T_{a-b}\rightarrow\infty$ for $\Delta\rightarrow-\infty$.  The limit
$t_c(b)$ is an {\it increasing} function of $|b|$, with $t_c(0)=3/(4\ln
2)\approx 1.08$.

For $v<0$, Eq.\ (\ref{e4r}) is to be replaced by
\begin{equation}
\Delta<1/2-t\ln[\frac{3(1+2e^{-3/2t})}{2(1-e^{-3/2t})^2}]\;\;\;(v<0)\,,
\label{e4rm}
\end{equation}
so that global bipartite entanglement starts for $\Delta<1/2$ ($\Delta<1/2-t\ln
(3/2)$ for $t\ll 1$), although for $t\gg 1$ Eq.\ (\ref{e4rm}) and (\ref{e4r})
are almost coincident, implying that $T_{a-bc}$ is almost the same as for $v>0$
for large negative $\Delta$ (Fig.\ \ref{f1}). This is not the case for
$T_{a-b}$, which for $v<0$ is determined by Eq.\ (\ref{e6r}) with
$\gamma=2+\alpha$, and is {\it lower} than the value for $v>0$, {\it vanishing}
for $b\rightarrow 0$ (no pairwise entanglement in the absence of field
\cite{WFS.01,GBF.03}). This effect is due to the larger degeneracy present for
$v<0$, where the ground state corresponds to $S=|M|=1/2$ for
$\Delta<\Delta_c^-(b)=1/2-|v/b|$, being then {\it two-fold} degenerate for
$b\neq 0$, and to an aligned state ($|M|=3/2$) if $\Delta>\Delta_c^-(b)$. For
$\Delta<\Delta_c^-(b)$ and $b\neq 0$, the mixture $\rho=\half P_{1/2,\pm 1/2}$
(the $T\rightarrow 0$ limit of $\rho(T)$) is still fully entangled, with
$N_{a-bc}=\sqrt{2}/6\approx 0.23$ and $N_{a-b}=(\sqrt{2}-1)/6\approx 0.07$.
However, for $b=0$, states $|S,\pm M\rangle$ become degenerate and the ensuing
$T\rightarrow 0$ limit, $\rho=\frac{1}{4}(P_{1/2,1/2}+P_{1/2,-1/2})$, has still
{\it global} entanglement ($N_{a-bc}=1/6$), but {\it  no pairwise entanglement}
($N_{a-b}=0$), explaining the vanishing of $T_{a-b}$ in this case.

\begin{figure}[t]
\vspace*{-1.5cm}

 \centerline{\hspace*{-1.5cm}\scalebox{0.75}{\includegraphics{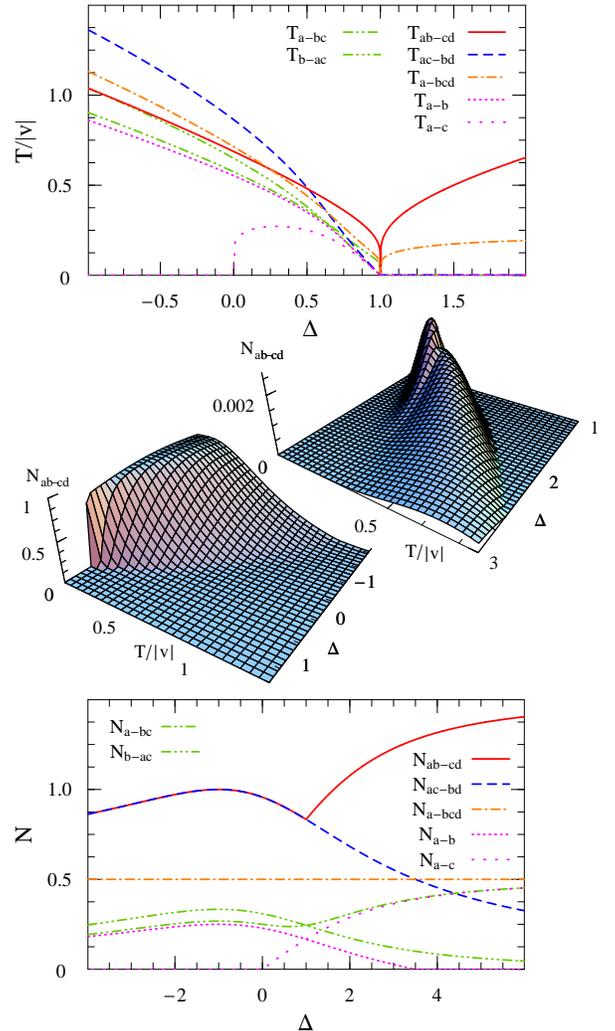}}}
 \vspace*{-7.cm}

\caption{(Color online). Top: Limit temperatures for global ($T_{ab-cd}$,
$T_{ac-bd}$, $T_{a-bcd}$), pairwise ($T_{a-b}$, $T_{a-c}$) and three-qubit
($T_{a-bc}$, $T_{b-ac}$) negativities for $n=4$ qubits in a nearest neighbor
cyclic chain, as a function of the asymmetry $\Delta$.  There is global
entanglement also for $\Delta>1$. Results for reduced systems correspond to
zero field. Center: The behavior of the global negativity $N_{ab-cd}$ as a
function of temperature and $\Delta$ (with different scales for $\Delta<1$ and
$\Delta>1$) for $b=0$. Bottom: Negativities of the state (\ref{psi0}) as a
function of $\Delta$. For $b=0$, they represent the ground state negativities
for $\Delta<1$.} \label{f2}\vspace*{-.25cm}
\end{figure}

{\it Four qubit case}. A surprise comes already for the $n=4$ chain (Fig.\
\ref{f2}), where for $T>0$, {\it global bipartite entanglement is seen to arise
also for $\Delta>1$}. The ground state is entangled just for
$\Delta<\Delta_c(b)$ [Eq.\ (\ref{dc})], with the transitions $|M|\rightarrow
|M|+1$ occurring at $\Delta=(1-2\bar{b}-\bar{b}^2)/(1+\bar{b})$,
$\bar{b}=|b/v|$ ($0\rightarrow 1$), and $\Delta=\Delta_c(b)$ ($1\rightarrow
2$), collapsing both into a single $0\rightarrow 2$ transition  at $\Delta=1$
for $b=0$. The corresponding limit temperatures are depicted in the top panel,
where it is seen that they all {\it vanish} for $\Delta\rightarrow 1$, but
those corresponding to $N_{ab-cd}$ (entanglement between {\it contiguous}
pairs) and $N_{a-bcd}$ (entanglement of one qubit with the rest) become again
non-zero for $\Delta>1$, indicating the reentry of the corresponding
entanglement. The highest limit temperature for $\Delta\alt 1/2$ corresponds to
$N_{ac-bd}$ (entanglement between {\it non-contiguous} pairs), but changes to
$N_{ab-cd}$ for $\Delta\agt 1/2$. The thermal behavior of $N_{ab-cd}$ for $b=0$
is depicted in the central panel, where it is seen that the reentry for
$\Delta>1$ is actually weak and vanishes smoothly for $T\rightarrow 0$. Let us
remark that no entanglement reentry for $\Delta>1$ takes place in the analogous
fully connected four qubit case.

The $M=0$ ground state in the four qubit cyclic chain is an entangled state of
the form
\begin{equation}|\Psi\rangle=
\alpha(|\!\uparrow\uparrow\downarrow\downarrow\rangle+\ldots)+
\beta(|\!\uparrow\downarrow\uparrow\downarrow\rangle+\ldots)\,,\label{psi0}
\end{equation}
where $\ldots$ denotes similar states obtained by translation and
$\beta/\alpha=(\sqrt{8+\Delta^2}-\Delta)/2$, with $4\alpha^2+2\beta^2=1$ and
$\beta>\alpha$ for $\Delta<1$. Its energy is $-|v|\beta/\alpha$. This state
exerts a strong influence on the entanglement of $\rho(T)$ even for $\Delta>1$.
It has maximum entanglement between one qubit with the rest, $N_{a-bcd}=1/2$
(the reduced one-qubit density $\rho_a$ is maximally mixed), while
$N_{ac-bd}=\beta(4\alpha+\beta)$, and $N_{ab-cd}=N_{ac-bd}$ if $\Delta<1$ and
$6\alpha^2-\beta^2$ if $\Delta>1$ (Fig.\ \ref{f2}, lower panel), becoming
$N_{ab-cd}$ much larger than $N_{ac-bd}$ for large $\Delta$. This seems to
affect the most persistent negativity of $\rho(T)$ for $\Delta>1$, allowing a
positive value of $N_{ab-cd}$ and reducing $N_{ac-bd}$ to 0. The $|M|=1$ ground
states are $W$-states
($\propto|\!\uparrow\downarrow\downarrow\downarrow\rangle+
{|\!\downarrow\uparrow\downarrow\downarrow\rangle} +\ldots$ for $M=-1$), with
lower values $N_{ab-cd}=N_{ac-bd}=1/2$ and $N_{a-bcd}=\sqrt{3}/4$.

We also depict in the upper panel the limit temperatures for $b=0$ of all
reduced negativities, namely $T_{a-bc}$, $T_{b-ac}$ for the three qubit density
$\rho_{abc}$ and $T_{a-b}$, $T_{a-c}$ for the reduced pair densities
$\rho_{ab}$ and $\rho_{ac}$. They all vanish for $\Delta>1$, so that in this
region just {\it global} bipartite entanglement persists (a result valid for
any $b$). It is also seen that while $T_{a-b}$ (adjacent qubits) increases as
$\Delta$ decreases, {\it diverging} for $\Delta\rightarrow-\infty$ (in contrast
with the behavior for $n=3$), $T_{a-c}$ (non-adjacent qubits) {\it vanishes}
for $\Delta<0$.  This is a consequence of the ground state behavior of
$N_{a-c}$, which vanishes for $\Delta<0$ (see bottom panel). The state
(\ref{psi0}) leads to $N_{a-b}=\alpha(2\beta-\alpha)$ for $\Delta<7/2$ (and $0$
for $\Delta>7/2$), and $N_{a-c}=2\alpha^2-\beta^2$ for $\Delta>0$ (and $0$ for
$\Delta<0$), with $N_{a-b}>N_{a-c}$ for $\Delta<1$. The ordering of
negativities is of course in agreement with the discussion of sec.\ II.

\begin{figure}[t]
\vspace*{-1.2cm}

 \centerline{\hspace*{-1.5cm}\scalebox{0.65}{\includegraphics{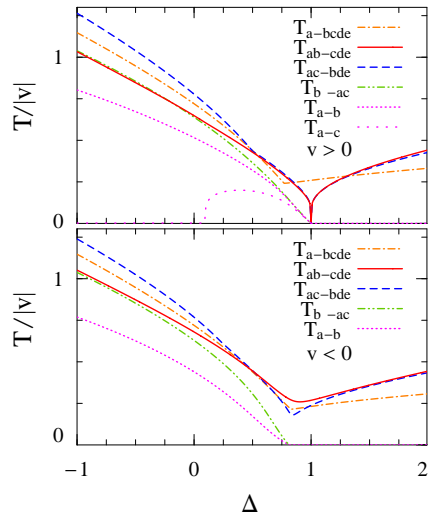}}}
 \vspace*{-11.25cm}

\caption{(Color online). The set of limit temperatures for global negativities,
$T_{a-bcde}$, $T_{ab-cde}$, $T_{ac-bde}$, for $n=5$ qubits in the cyclic chain
as a function of $\Delta$ for $v>0$ (top) and $v<0$ (bottom). The concomitant
limit temperatures for pairwise ($T_{a-b}$, $T_{a-c}$) and the most persistent
three-qubit ($T_{b-ac}$) negativity for $b=0$ are also depicted. $T_{a-c}$
vanishes for $v<0$ and $b=0$.}
 \label{f3}\vspace*{-.25cm}
\end{figure}

The behavior of global negativities for $n=5$ qubits and $v>0$ (Fig.\ \ref{f3},
upper panel) is quite similar, although there are some important changes: a)
For $T>0$, {\it all} global negativities exhibit a reentry for $\Delta>1$,
including that of the non-contiguous $2-3$ partition, $N_{ac-bde}$, with
$T_{ac-bde}$ lying close to $T_{ab-cde}$ (contiguous 2-3 partition) for
$\Delta\agt 1/2$; b) The negativity of one-qubit with the rest, $N_{a-bcde}$,
remains non-zero $\forall$ $\Delta$ if $T>0$, {\it including} the isotropic
case $\Delta=1$, providing the highest limit temperature for $0.85\alt
\Delta\alt 1.35$; c) For $\Delta>1$ there is no pairwise nor three qubit
entanglement at any field, but there is four qubit entanglement (all reduced
four-qubit negativities are small but non-zero). Besides, for
$\Delta\rightarrow-\infty$ the limit temperature for adjacent pairwise
entanglement $T_{a-b}$ approaches a finite limit \cite{GBF.03}, as occurs for
$n=3$, exhibiting a maximum at $\Delta\approx -5.6$ for $b=0$, but all those
for three and four adjacent qubits ($T_{a-bc}$, $T_{b-ac}$, $T_{a-bcd}$,
$\ldots$) diverge, as the global ones. Note also that $T_{a-c}$ vanishes below
a certain limit ($\Delta\alt 0.1$ for $b=0$), reflecting the vanishing of
$N_{a-c}$ at $T=0$ below this value. For $n=5$ and $v>0$, the ground state
transitions $|M|\rightarrow|M|+1$ are located at
$\Delta=(1-\bar{b}-\bar{b}^2)/(1+\bar{b})$
($\frac{1}{2}\rightarrow\frac{3}{2}$) and $\Delta_c(b)$
($\frac{3}{2}\rightarrow \frac{5}{2}$), collapsing into a single
$\frac{1}{2}\rightarrow \frac{5}{2}$ transition at $\Delta=1$ if $b=0$.

Global limit temperatures for $n=5$ and $v<0$ (lower panel) are quite close to
the previous ones. The main difference is that {\it all} of them remain
non-zero for all $\Delta$, the most persistent negativity corresponding to
$N_{ac-bde}$ if $\Delta\alt 0.5$ and  $N_{ab-cde}$ if $\Delta\agt 0.5$. Limit
temperatures for pairwise entanglement exhibit more significant differences.
For $b=0$ there is no non-contiguous pairwise entanglement if $v<0$
\cite{GBF.03} ($T_{a-c}=0$), since the ground state is four-fold degenerate and
leads to $N_{a-c}=0$ $\forall$ $\Delta$, while $T_{a-b}$ is lower (saturating
again for $\Delta\rightarrow-\infty$). For $v<0$, the ground state transitions
$|M|\rightarrow |M|+1$ occur at
$\Delta=-\frac{(\bar{b}-1/2)(4\bar{b}+3+\sqrt{5})}{4\bar{b}+1+\sqrt{5}}$ and
$\Delta=\Delta_c^-(b)=\frac{1}{4}(1+\sqrt{5})-\bar{b}$, with both collapsing at
$\Delta_c^-(0)\approx 0.81$ for $b=0$. Again, for $\Delta>\Delta_c(0)$ there is
no two- nor three-qubit negativity, but there is four qubit entanglement.

\begin{figure}[t]
\vspace*{0.5cm}

 \centerline{\hspace*{-1.5cm}\scalebox{0.65}{\includegraphics{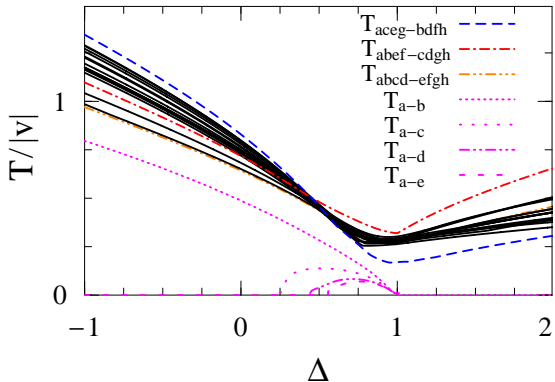}}}
 \vspace*{0.cm}

\caption{(Color online). The limit temperatures of all seventeen global
negativities for $n=8$ qubits in a cyclic chain as a function of asymmetry. The
highest and lowest temperatures in each sector are specially indicated. The
limit temperatures for adjacent and non-adjacent pairwise entanglement for
$b=0$ are also depicted.}
 \label{f4}\vspace*{-0.25cm}
\end{figure}

In order to appreciate the trend for larger $n$, Fig.\ \ref{f4} depicts the set
of limit temperatures of the 17 distinct global negativities existing for $n=8$
qubits $ab\ldots gh$, consisting of $N_{a-*}$ (one qubit with the rest), and
those between different pairs and the rest ($N_{ab-*}$, $N_{ac-*}$, $N_{ad-*}$,
$N_{ae-*}$), three qubits and the rest ($N_{abc-*}$, $N_{abd-*}$, $N_{abe-*}$,
$N_{ace-*}$, $N_{acf-*}$) and four qubits and the rest ($N_{abcd-*}$,
$N_{abce-*}$, $N_{abcf-*}=N_{abde-*}$, $N_{abef-*}$, $N_{abdf-*}$,
$N_{abdg-*}$, $N_{aceg-*}$). {\it All} these temperatures remain non-zero
$\forall\Delta$, including the region $\Delta\geq 1$ where the ground state is
aligned for any $b$, confirming the tendency observed for $n=4$ and $5$. The
highest limit temperature corresponds to the full non-contiguous 4-4 partition
($T_{aceg-*}$) for $\Delta\alt 0.5$, but changes to that between non-contiguous
pairs of adjacent qubits ($T_{abef-*}$) for $\Delta\agt 0.5$ (in this region
$T_{aceg-*}$ becomes the lowest global limit temperature). Again, the ``depth''
of the entanglement for $\Delta>1$ is up to four qubit subsystems, being zero
all pairwise and three-qubit negativities. The limit temperature for adjacent
pairwise entanglement for $b=0$ lies well below the bundle of global limit
temperatures, while those of non-adjacent pairs are small and non-zero just in
a small interval between $\Delta=0$ and $\Delta=1$, where the ground state
negativity is non-zero. Let us also remark that the revival for $\Delta>1$ is
sensitive to the interaction range. For instance, there is no entanglement
revival in the analogous fully connected case.

\begin{figure}
\vspace*{-1.25cm}

 \centerline{\hspace*{-1.5cm}\scalebox{0.65}{\includegraphics{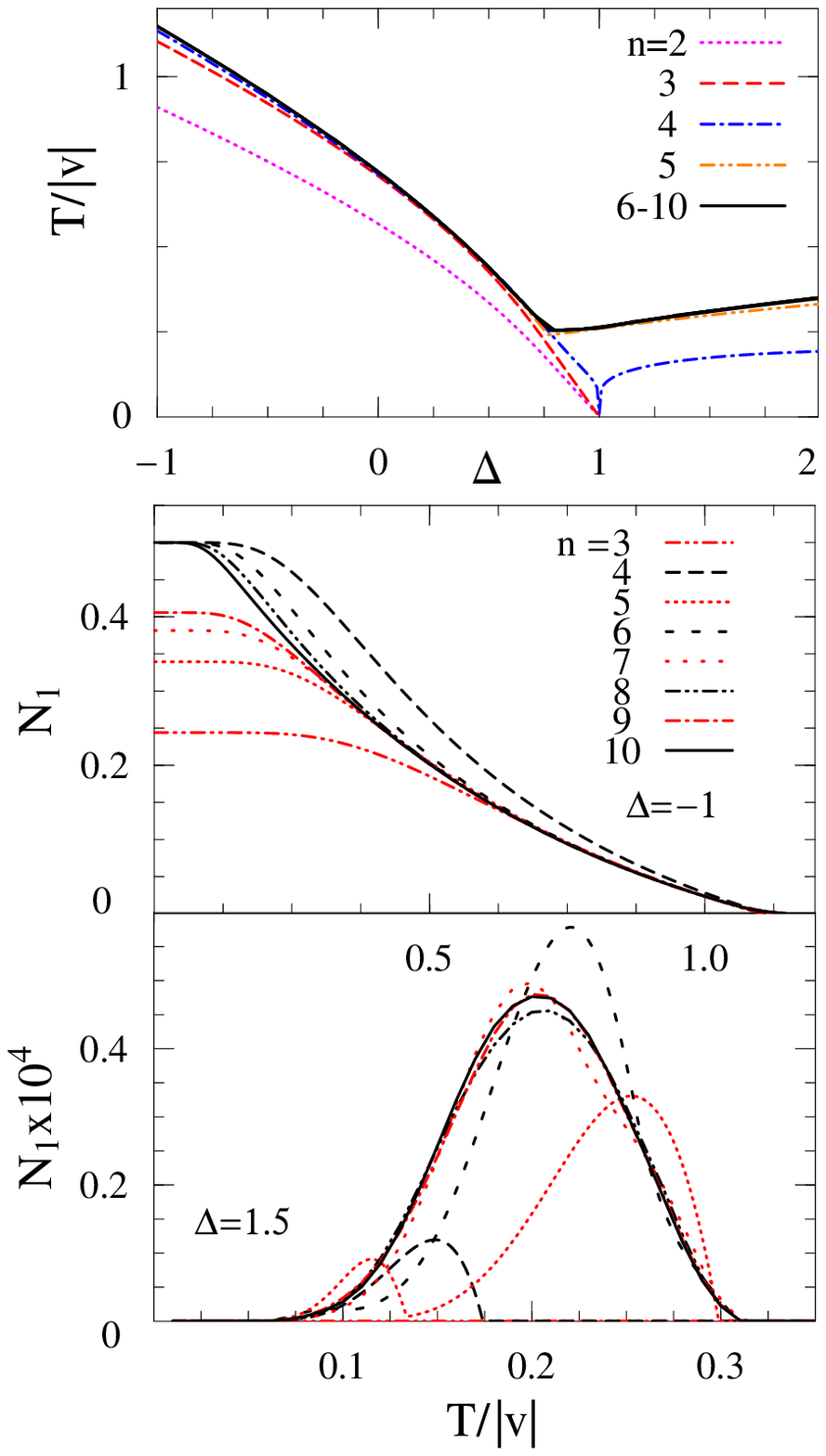}}}
 \vspace*{-7.5cm}

\caption{(Color online) Top: Limit temperatures for non-zero negativity between
one qubit and the remaining $n-1$ qubits, for $v>0$ and $n=2,\ldots 10$. Center
and bottom: The thermal behavior of these negativities for $\Delta=-1$ and
$1.5$, for $b=0$.}
 \label{f5}\vspace*{-.25cm}
\end{figure}

Finally, Fig.\ (\ref{f5}) compares results for the entanglement between one
qubit with the rest for different $n$, assuming $v>0$. As previously stated,
$N_{a-*}$ and hence $T_{a-*}$ are upper bounds to all pairwise negativities and
limit temperatures. It is first seen in the upper panel that the global limit
temperatures $T_{a-*}$ for different $n$ rapidly converge to a common value,
results for $n=6-10$ being undistinguishable in the scale and range of the
figure. This temperature exhibits a slope discontinuity around $\Delta\approx
0.8$, where the number $k$ of negative eigenvalues of the partial transpose
exhibits a minimum (while in the case of pure states $k=1$ for $N_{a-*}$ [Eq.\
(\ref{N0})], it can be much larger than 1 for non-pure states).

The corresponding global negativities $N_{a-*}$ also approach a limit curve for
$\Delta<1$, as seen in the central panel for $b=0$ and $\Delta=-1$. Note that
for {\it all} even $n$, $N_{a-*}$ starts at its maximum value $1/2$ at $T=0$,
since the reduced one-qubit density $\rho_a$ for the $M=0$ ground state is {\it
maximally mixed} and Eq.\ (\ref{N0}) reaches then its maximum value. At fixed
$T>0$, they tend initially to decrease for increasing $n$, rapidly approaching
the limit curve. On the other hand, for $n$ odd and $b=0$, the ground state for
$\Delta=-1$ has $M=\pm 1/2$ and is degenerate, so that the initial value is
lower than in the even case (see the comment for $n=3$), although it increases
as $n$ increases and approaches the limit curve (for $b\neq 0$, $N_{a-*}$ would
start at $\approx 0.47$ for $n=3$, increasing then with $n$). The lower panel
depicts the reentry of the negativity for $T>0$ at $\Delta=1.5$ for increasing
$n$. The behavior is less monotonous than in the previous case, although the
difference between results for neighboring odd-even values of $n$ tend to
decrease as $n$ increases.

\section{Conclusions}
The present results demonstrate that limit temperatures for global bipartite
entanglement in $XXZ$ chains may differ considerably from those limiting
pairwise entanglement between two qubits in the same chain. In contrast with
the latter, global limit temperatures are independent of the field $b$ and do
not saturate for $\Delta\rightarrow-\infty$ in odd chains. Moreover, global
thermal entanglement in nearest-neighbor cyclic chains may naturally arise at
low levels even for $\Delta>1$ if $n\geq 4$ (and $\Delta\geq 1$ if $n\geq 5$),
i.e., in anisotropic and isotropic chains with fully separable ground states,
generating entangled bipartitions of the whole system as well as of subsystems
with no less than four qubits. The ordering of global limit temperatures also
changes as $\Delta$ increases from negative to large positive values. Many of
our results rapidly saturate as $n$ increases (Fig.\ \ref{f5}), indicating that
they may remain stable for chains with a larger number of qubits. The present
study provides thus a more comprehensive understanding of the borders of
thermal entanglement in $XXZ$ spin chains, suggesting that quantum schemes
based on entanglement between multi-spin parties may be more resistant to the
effects of thermal randomness.

{\it Acknowledgment.} NC and RR acknowledge support, respectively, from CONICET
and CIC of Argentina.

\subsection*{Appendix}
{\it Negativity of pure states}. Eq.\ (\ref{N0}) can be rewritten as
\[N_p[\rho]=\half\{[S_f(\rho_{\{m\}})+1]^2-1\}\,, \]
where
\[S_f(\rho_{\{m\}})={\rm Tr}f(\rho_{\{m\}})\,,
\;\;f[\rho]\equiv\sqrt{\rho}-\rho\,, \] is a simple non-additive entropy of
$\rho_{\{m\}}$ \cite{RC.02}, since the function $f$ is concave in $[0,1]$ and
satisfies $f(0)=f(1)=0$. As a consequence, $S_f(\rho_{\{m\}})\geq 0$, being
maximum for $\rho_{\{m\}}=I/d_m$ and zero when $\rho_{\{m\}}$ is pure, and
satisfies $S_f(\rho_{\{m\}})\geq S_f(\rho'_{\{m\}})$ if
$\rho_{\{m\}}\prec\rho'_{\{m\}}$, where the last condition means that the
eigenvalues of $\rho_{\{m\}}$ are majorized by those of $\rho'_{\{m\}}$
\cite{RC.02}. These properties are then also satisfied by the pure state
negativity $N_p$ since it is an increasing function of $S_f$ that vanishes for
$S_f=0$, and ensure that it cannot increase under LOCC.

{\it Independence of global limit temperatures from the uniform magnetic field
$b$}. Let us consider a Hamiltonian
\[H=bS_z+V\,,\]
where $V$ is independent of $b$ and satisfies $[V,S_z]=0$. For an arbitrary
bipartition $p=\{m\}-\{n-m\}$ of an $n$-qubit system and a standard basis of
states $|M_{\{m\}},M-M_{\{m\}}\rangle$, where $M$ is the eigenvalue of the
total spin component $S_z$ and $M_{\{m\}}$ that for the first subsystem
(remaining labels omitted), the partial transpose of the exponential
$D(b)=\exp[-H/T]=\exp[-bS_z/T]\exp[-V/T]$ in the previous basis satisfies
\cite{RC.05}
\[D^{t_p}(b)=\exp[-bS_z/2T]D^{t_p}(0)\exp[-bS_z/2T],\]
where $D^{t_p}(0)$ has matrix elements between states with total spin component
$M$ and $M\pm 2k$, with $k$ integer. While the negativity will depend on $b$,
the limits for the positivity of $D^{t_p}(b)$ will not, since $\exp[-bS_z/2T]$
is real and positive, being the same as those for $D^{t_p}(0)$. Global limit
temperatures will then be independent of $b$. This applies in particular to any
$XXZ$ type Hamiltonian.

{\it Negativity of mixtures of two- and three-qubit states with good angular
momentum.} The negativity $N_{a-b}$ of any two qubit mixed state of the form
\begin{equation}
\rho=\sum_{S=0,1}\sum_{M=-S}^Sp^S_M|SM\rangle\langle SM|\,,\label{r1}
 \end{equation}
where $|SM\rangle$ denote the states with good total spin $S$ and component
$M$, can be shown to be \cite{CR.04}
\begin{equation}
N_{a-b}=\half{\rm Max}
[\sqrt{(p^1_{0}-p^0_{0})^2+(p^1_{1}-p^1_{-1})^2}-p^1_1-p^1_{-1},0]\label{N2}
\end{equation}
The condition $N_{a-b}>0$ leads then to Eq.\ (\ref{EP}). Note also that the
single qubit density is $\rho_a={\rm Tr}_b\rho=\sum_{\nu=\pm
1/2}q_\nu|\nu\rangle\langle \nu|$, with $q_\nu=p^1_\nu+\half(p^0_0+p^1_0)$.

For $n=3$ qubits, a set of 8 orthonormal states with good total spin $S$ and
spin component $M$ is given by
$|\frac{3}{2}\frac{3}{2}\rangle=|\!\uparrow\uparrow\uparrow\rangle$,
$|\frac{3}{2}\frac{1}{2}\rangle=(|\!\downarrow\uparrow\uparrow\rangle
+|\!\uparrow\downarrow\uparrow\rangle+
|\!\downarrow\downarrow\uparrow\rangle)/\sqrt{3}$,
$|\frac{1}{2}\frac{1}{2}a\rangle= (|\!\downarrow\uparrow\uparrow\rangle-
|\!\uparrow\downarrow\uparrow\rangle)/\sqrt{2}$,
$|\frac{1}{2}\frac{1}{2}b\rangle=(|\!\downarrow\uparrow\uparrow\rangle+
|\!\uparrow\downarrow\uparrow\rangle-2|\!\uparrow\uparrow\downarrow\rangle)
/\sqrt{6}$, and the corresponding partners for $M\rightarrow -M$. For a
three-qubit density of the form
\begin{equation}
\rho=\sum_{S=1/2,3/2}\sum_{M=-S}^{S}p^S_MP^S_M\,,\label{r2}
\end{equation}
where $P^S_M$ denotes the projector onto the subspace with total spin $S$ and
spin component $M$, an analytic evaluation of the eigenvalues of the partial
transpose of $\rho$ yields the following expression for the global negativity
$N_{a-bc}=N_{ab-c}=N_{ca-b}$,
\begin{eqnarray}
&&N_{a-bc}={\textstyle\frac{1}{6}} \sum_{\nu=\pm 1}{\rm Max}[\nonumber\\
&&\sqrt{(3p^{3/2}_{-3\nu/2}-2p^{3/2}_{\nu/2}-
p^{1/2}_{\nu/2})^2+8(p^{3/2}_{-\nu/2}-p^{1/2}_{-\nu/2})^2}\nonumber\\
&&-3p^{3/2}_{-3\nu/2}
-2p^{3/2}_{\nu/2}-p^{1/2}_{\nu/2}),0]\,.\label{A1}
\end{eqnarray}
The condition $N_{a-bc}>0$ leads then to Eq.\ (\ref{e3r}). The ensuing reduced
two qubit density $\rho_{a-b}={\rm Tr}_c\rho$ (identical to $\rho_{a-c}$,
$\rho_{b-c}$) has again the form (\ref{r1}), with
\begin{eqnarray}
p^1_{\pm 1}&=&(3p^{3/2}_{\pm 3/2}+p^{3/2}_{\pm 1/2}+2p^{1/2}_{\pm 1/2})/3,
\nonumber\\
p^1_{0}&=&(2(p^{3/2}_{1/2}+p^{3/2}_{-1/2})+p^0_0)/3,\;\;p^0_0=p^{1/2}_{1/2}
+p^{1/2}_{-1/2}\nonumber
\end{eqnarray}
The condition (\ref{EP}) for pairwise entanglement leads then to
\begin{equation}
 |\sum_{\nu=\pm 1}p^{3/2}_{\nu/2}-p^{1/2}_{\nu/2}|>\prod_{\nu=\pm 1}
\sqrt{3p^{3/2}_{3\nu/2}+p^{3/2}_{\nu/2}+2p^{1/2}_{\nu/2}}
\label{e5r}\,.
\end{equation}
In the thermal case for the Hamiltonian (\ref{H1}), Eq.\ (\ref{e5r}) leads to
Eq.\ (\ref{e6r}).

\end{document}